\documentclass[preprint,showpacs,preprintnumbers,amsmath,amssymb]{revtex4}
\usepackage{amssymb}
\usepackage{amsmath}
\usepackage{graphicx}

\begin{document}

%
%
%
\title{Electromagnetic Pulse Driven Spin-dependent Currents in Semiconductor
Quantum Rings}

\author{Zhen-Gang Zhu, and Jamal Berakdar}
\affiliation{ Institut f\"ur Physik, Martin-Luther Universit\"at
Halle-Wittenberg, Heinrich-Damerow-Str.4 06120 Halle, Germany}

\begin{abstract}
We investigate the non-equilibrium charge  and spin-dependent currents in a
quantum ring with a Rashba spin orbit interaction (SOI) driven by
two asymmetric picosecond electromagnetic  pulses. The equilibrium
persistent charge  and persistent spin-dependent currents are investigated as
well. It is shown that the dynamical charge and the dynamical spin-dependent currents vary smoothly with a static external magnetic flux and the
SOI provides a SU(2) effective flux that changes the phases of the
dynamic charge  and the dynamic spin-dependent currents. The period of the
oscillation of the total charge current with the delay time between the  pulses
is larger in a quantum ring with a larger radius. The
parameters of the pulse fields control to a certain extent the total
charge  and the total spin-dependent currents. The calculations are applicable
to nano-meter rings fabricated in heterojuctions of III-V and II-VI
semiconductors containing several hundreds electrons.
\end{abstract}

\pacs{78.67.-n, 71.70.Ej, 42.65.Re, 72.25.Fe} \maketitle
%
%
%


%
%
%

\section{Introduction}
Study of the spin-orbit interaction (SOI) in semiconductor low
dimensional structures and its application for spintronics devices
have attracted much attention recently \cite{spintronics}. There are
two important kinds of SOI in conventional semiconductors: one is
the Dresselhaus SOI induced by bulk inversion asymmetry
\cite{dresselhaus}, and the other is the Rashba SOI caused by
structure inversion asymmetry \cite{rashba}. As pointed out in \cite{lommer},
the Rashba SOI is dominant in a narrow gap semiconductor system
and the strength of the Rashba SOI can be tuned by an
external gate voltage in HgTe \cite{hgte}, InAs \cite{luo},
In$_{x}$Ga$_{1-x}$As \cite{nitta}, and GaAs \cite{rashba,malcher}
quantum wells.  Recent
research is focused on the electrically-induced generation of a spin-dependent current (SC) mediated by SOI-type mechanism, e.g. as in the
intrinsic spin Hall effect in a 3D p-doped semiconductor
\cite{murakami} and in a 2D electron gas with Rashba SOI
\cite{sinova}. Here we study a high quality
spin-interacting quantum rings (QRs) with a radius on the nanometer
scale  \cite{fuhrer,ring}. These systems show Aharonov-Bohm-type
(AB) spin-interferences \cite{nitta99,nitta03}. In particular we
investigate  the dynamics triggered by time-dependent
electric fields as provided by time-asymmetric pulses \cite{hcp} 
or tailored laser pulse sequences  \cite{bennett}. The quantity
under study is the spin-resolved pulse-driven current, in analogy to
the spin-independent case \cite{matos05,pershin,rasanen}. In a
previous work \cite{zhu}, we investigated the dynamical response of
the charge polarization to the pulse application.  No net charge or
spin-dependent current is generated because the clock-wise and
anti-clock-wise symmetry of the carrier is not broken by one pulse
or a series of pulses
 having the same linear polarization axis.  This symmetry
 is lifted  if two time-delayed pulses with non-collinear polarization axes  are
 applied  \cite{matos05}. However, to our knowledge all previous
 studies on light-induced currents in quantum rings did not
 consider the coupling of the spin to the orbital motion (and hence to the light field), which
 is addressed in this work. As detailed below, having done that, it is
 possible  to control dynamically  the spin-dependent
current in a 1D quantum ring with Rashba SOI by using two
time-delayed linearly polarized electromagnetic pulses. For
transparent interpretation of the results only the Rashba SOI is
considered in this work. The presence of the Dresselhaus SOI may
change qualitatively the results presented here for the
spin-dependent non-equilibrium dynamic of the carriers, which can be
anticipated from the findings on for the equilibrium case
\cite{ref1}.

\section{Theoretical model}
We study the response of charges and spins confined in a
one-dimensional (1D) ballistic QR with SOI to the application of two
short time-delayed linearly polarized asymmetric electromagnetic
pulses \cite{matos05,matos}. The effective single particle
Hamiltonian reads
$\hat{H}'=\hat{H}_{\mbox{SOI}}+\hat{H}_{1}(t)$ \cite{zhu},
with
\begin{eqnarray}
\hat{H}_{\mbox{SOI}}&=&\frac{\mathbf{p}^{2}}{2m^{*}}+V(\mathbf{r})+\frac{\alpha_{R}}{\hbar}(\hat{\mathbf{\sigma}}\times\mathbf{p})_{z}
, 
\hat{H}_{1}(t)=-e\mathbf{r}\cdot\mathbf{E}(t)+\mu_{B}\mathbf{B}(t)\cdot\hat{\mathbf{\sigma}}.
\label{h1t}
\end{eqnarray}
  $\mathbf{E}(t)$ and $\mathbf{B}(t)$ are the electric
 and the magnetic fields   of the pulse. Integrating out the $r$ dependence
$\hat{H}_{\mbox{SOI}}$ reads in cylindrical coordinates
\cite{meijer,frustaglia,molnar,foldi,sheng}
\begin{equation}
\hat{H}_{\mbox{SOI}}=\frac{\hbar\omega_{0}}{2}[(i\partial_{\varphi}+\frac{\phi}{\phi_{0}}-\frac{\omega_{R}}{2\omega_{0}}\sigma_{r})^{2}-
(\frac{\omega_{R}}{2\omega_{0}})^{2}+\frac{\omega_{B}}{\omega_{0}}\sigma_{z}].
\label{hsoi1}
\end{equation}
 $\partial_{\varphi}=\frac{\partial}{\partial\varphi}$,
$\phi_{0}=h/e$ is the flux unit, $\phi=B\pi a^{2}$ is the magnetic
flux threading the ring, $a$ is the radius of the ring,
$\hbar\omega_{0}=\hbar^{2}/(m^{*}a^{2})=2E_{0}$,
$\hbar\omega_{R}=2\alpha_{R}/a$, $\hbar\omega_{B}=2\mu_{B}B$ and $B$
are due to  a possible  external static magnetic field
$\mathbf{B}=B\hat{\mathbf{e}}_{z}$.
%
The single-particle eigenstates of  $\hat{H}_{\mbox{SOI}}$ are
represented as
$\Psi_{n}^{S}(\varphi)=e^{i(n+1/2)\varphi}\nu^{S}(\gamma,\varphi)$
where
$\nu^{S}(\gamma,\varphi)=( a^{S}e^{-i\varphi/2},
b^{S}e^{i\varphi/2})^{T} 
$ 
are spinors in the angle dependent local frame,  and
$a^{\uparrow}=\cos(\gamma/2),\;
b^{\uparrow}=\sin(\gamma/2),\;
a^{\downarrow}=-\sin(\gamma/2), b^{\downarrow}=\cos(\gamma/2),$ (\textit{T} means
transposed) where
$\tan\gamma=-Q_{R}=-\omega_{R}/\omega_{0}$ (if we ignore the Zeeman
splitting caused by the static magnetic field \cite{sheng,splett}).
$\gamma$ describes the direction of the spin quantization axis, as
illustrated in Fig. (1a). The energy spectrum of the QR with the SOI
reads
 \cite{frustaglia,molnar,foldi,sheng,splett,zhu}
\begin{equation}
E_{n}^{S}=\frac{\hbar\omega_{0}}{2}\left[(n-\phi/\phi_{0}+\frac{1-Sw}{2})^{2}-\frac{Q_{R}^{2}}{4}\right],
\label{eigenenergy}
\end{equation}
\[w=\sqrt{1+Q_{R}^{2}}=1/\cos\gamma,\] where $S=+1\, (S=-1)$  stands for spin up
(spin down) in the  local frame.
\begin{figure}[tbph]
\centering \includegraphics[width =10 cm, height=9 cm]{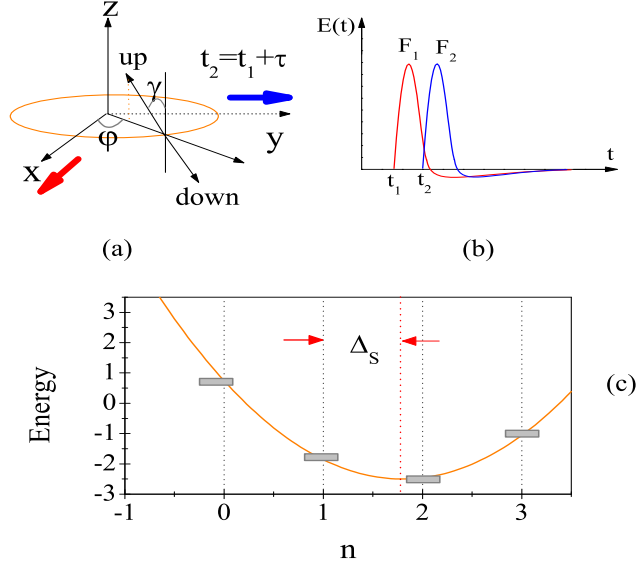}
\caption{(Color online) (a) Schematic graph of the geometry, spin
configuration and the applied pulses is shown. (b) Time-delayed
asymmetric pulses are schematically drawn. (c) Energy spectrum for a
ring with spin orbit interaction. $\Delta_{S}$ defines  the distance
between the spectrum symmetry axis and the smallest nearest
integer.} \label{fig1}
\end{figure}
%
\section{Pulse-driven single-particle dynamics}
We apply two time-asymmetric pulses to the system \textit{(see Fig.
(1b))}. The first one (at $t=0$) propagates in the $z$ direction and
has a duration $\tau_{d}$. Its E-field is along the $x$ direction.
$\tau_d$ is chosen much shorter than the ballistic time of the
carriers in which case the QR states develop as
\cite{matos05,matos,mizushima}
\begin{equation}
\Psi_{n}^{S}(\varphi,t>0)=\Psi_{n}^{S}(\varphi,t<0)e^{i\alpha_{1}\cos\varphi},
\; \alpha_{1}=eap/\hbar,\; p=-\int_{0}^{\tau_{d}}E(t)dt,
\label{matching}
\end{equation}
where $E(t)=Ff(t)$, $F$ and $f(t)$ describe the amplitude and the
time dependence of the electric field of the pulse respectively. In
the following, we use $F_{1}$ and $F_{2}$ to characterize the first
and the second pulses. The pulse effect is encapsulated entirely in
the \emph{action parameter} $\alpha_{1}$. With the initial
conditions $n(t<0)=n_{0}$ and $S(t<0)=S_{0}$  and using Eq.
(\ref{matching}) one finds
\begin{equation}
\Psi_{n_{0}}^{S_{0}}(\varphi,t)=\frac{1}{\sqrt{2\pi}}
\sum_{ns}C_{n}^{S}(n_{0},S_{0},t)e^{i(n+1/2)\varphi}e^{-iE_{n}^{S}t/\hbar}|\nu^{S}\rangle,
\label{wavef1}
\end{equation}
with
\begin{equation}
C_{n}^{S}=\left\{
\begin{array}{l l}
\delta_{SS_{0}}\delta_{nn_{0}} & \mbox{for t}\leq0,\\
\delta_{SS_{0}}i^{n_{0}-n}J_{n_{0}-n}(\alpha_{1}) & \mbox{for t}>0,
\end{array}
\right. \label{coeff}
\end{equation}
where $J_{n}$ is the n-th order Bessel function. For the
time-dependent energy we find
\begin{equation}
E_{n_{0}}^{S_{0}}(t>0)=E_{n_{0}}^{S_{0}}(t<0)+
\frac{\hbar\omega_{0}}{2}\frac{\alpha^{2}_{1}}{2}, \label{energy3}
\end{equation}
with $E_{n_{0}}^{S_{0}}(t<0)$ is given by Eq. (\ref{eigenenergy}).
Applying a second pulse at $t=\tau$ with the same duration
$\tau_{d}$ but the electric field being along the $y$ axis
\textit{(see Fig. (1b))}, the wave functions develop as
$\Psi_{n_{0}}^{S_{0}}(\varphi,t>\tau)=\Psi_{n_{0}}^{S_{0}}(\varphi,t<\tau)e^{i\alpha_{2}\sin\varphi}$,
where $\alpha_{2}$ is the action parameter associated with the
second pulse. $\Psi_{n_{0}}^{S_{0}}(\varphi,t=\tau^{-})$ follows
from Eq. (\ref{wavef1}). For $t>\tau$ the expansion coefficients
behave as
$C_{n'}^{S'}(n_{0},S_{0},t>\tau)=\sum_{n}\delta_{S'S_{0}}[i^{n_{0}-n}J_{n_{0}-n}(\alpha_{1})
\; J_{n'-n}(\alpha_{2})]
e^{i(E_{n'}^{S'}-E_{n}^{S_{0}})\tau/\hbar}.$
\section{Nonequilibrium spin and charge currents}
A single pulse  does not generate in QR any net charge current
because of the degeneracy of the orbital states. However, the charge
will be polarized \cite{matos,zhu} and corresponding dipole moments
oscillate in the $x$ direction with an associated  optical emission.
Applying a  second pulse as described above leads to a
non-equilibrium net current, in addition to the persistent charge
current caused by the static flux and the SOI which causes   a SU(2)
vector potential and manifests itself in an induced spin-dependent
persistent charge current \cite{sheng,splett,oh,frohlich}.
Consequently, a non-equilibrium spin-dependent current is induced.

The line velocity operator is \cite{ballentine}
\[ \hat{\mathbf{v}}_{\varphi}=\hat{\mathbf{e}}_{\varphi}
\left\{\frac{-i\hbar}{m^{*}a}\partial_{\varphi}-\frac{\hbar}{m^{*}a}\frac{\phi}{\phi_{0}}
+\frac{\alpha_{R}}{\hbar}\sigma_{r}\right\}\] which is associated to the
operator of the angular velocity $\hat{\mathbf{v}_{\varphi}}/a$
\cite{loss}. Contributions to the persistent charge current from
each QR  level read \cite{wendler}
\[
\mathbf{I}_{n_{0},S_{0}}=
\frac{1}{2\pi}\int_{0}^{2\pi}d\varphi\int_{r_{1}}^{r_{2}}dr
\mathbf{j}^{\varphi}_{n_{0},S_{0}}(\mathbf{r}',t>\tau),\]
where
\[\mathbf{j}^{\varphi}_{n_{0},S_{0}}=e\Re[\Psi_{n_{0}}^{S_{0},\dag}
(\mathbf{r}',t)\hat{\mathbf{v}}_{\varphi}
\Psi_{n_{0}}^{S_{0}}(\mathbf{r}',t)].\]
 Upon algebraic manipulations
we find
\begin{equation}
\mathbf{I}_{n_{0},S_{0}}(t>\tau)=\mathbf{I}_{n_{0},S_{0}}^{(0)}(t>\tau)+\mathbf{I}_{n_{0},S_{0}}^{(1)}(t>\tau).
\label{pccall}
\end{equation}
The index ``(0)" stands for the static persistent charge current (PCC)
which exists in the absence of pulse field, whereas the index ``(1)"
indicates the pulse-induced dynamic charge current (DCC). The PCC is
caused by a magnetic U(1) flux and has been studied extensively
\cite{pccreview} without \cite{wendler,buttiker,chakraborty} or with
the spin interactions \cite{loss}. It has been experimentally
observed both in gold rings of radius with 1.2 and 2.0 $\mu$m
\cite{chandrasekhar} and in a GaAs-AlGaAs ring of radius about 1
$\mu$m \cite{mailly}. The SOI scattering effects  were also studied
\cite{sopcc}. The PCC carried by the states characterized by $n_{0}$
and $S_{0}$ reads (please note the current in this work is defined
as flow of positive charges, which is opposite to the direction of
flow of electrons)
\begin{equation}
\mathbf{I}_{n_{0},S_{0}}^{(0)}(t>\tau)=
\hat{\mathbf{e}}_{\varphi}I_{0}\left(n_{0}-\frac{\phi}{\phi_{0}}+\frac{1-S_{0}w}{2}\right),
\label{pcc0}
\end{equation}
where $I_{0}=2E_{0}a/\phi_{0}$ is the unit of CC, the second term on
the right hand side of Eq. (\ref{pcc0}) stems from the
static magnetic field; the third term is a consequence of the
SU(2) flux of the SOI \cite{sheng}. The DCC part is
\begin{equation}
\mathbf{I}_{n_{0},S_{0}}^{(1)}(t>\tau)=\hat{\mathbf{e}}_{\varphi}I_{0}
\left\{\alpha_{2}\langle\cos\varphi\rangle_{n_{0}}^{S_{0}}(\tau)\right\},
\label{pcc1}
\end{equation}
where
\[\langle\cos\varphi\rangle_{n_{0}}^{S_{0}}(\tau)=\alpha_{1}h(\Omega_{\tau1})\sin
b_{\tau}\cos[2(n_{0}-\frac{\phi}{\phi_{0}}+\frac{1-S_{0}w}{2})b_{\tau}],
\] \[b_{\tau}=\omega_{0}\tau/2,\]
\[\Omega_{\tau1}=\alpha_{1}\sqrt{2(1-\cos(2b_{\tau}))},\]
\[h(\Omega_{\tau1})=J_{0}(\Omega_{\tau1})+J_{2}(\Omega_{\tau1}).\]

To obtain the total persistent charge current and  the dynamic
current  we have to consider the spin-resolved occupations of the
single particle states. For simplicity we operate at zero
temperatures and ignore the relaxation caused by phonons or other
mechanisms, i.e. we confine ourself to times  shorter than the
relaxation time. The general case can be developed along the line of
Ref. \cite{moskalenko}.

At first we introduce an effective flux as
\begin{equation}
\phi_{S}=\phi-\phi_{0}\frac{1-Sw}{2}. \label{effflux}
\end{equation}
As evident from Eq. (\ref{eigenenergy}) the spectrum is symmetric
with respect to $x_{S}=\phi_{S}/\phi_{0}$. Further we define  the
shift $\Delta_{S}=x_{S}-l(l')$, where $S=\uparrow$ or $\downarrow$.
Here $l(l')=[x_{\uparrow(\downarrow)}]$ where $[x]$ means the
nearest integer which is less than $x$. $\Delta_{S}$ is shown in
Fig. 1. When $\Delta_{S}=1$, it is equivalent to $\Delta_{S}=0$.
Furthermore, $\bar{\Delta}_{S}=|1/2-\Delta_{S}|$ is the distance
between the $x_{S}$ and the nearest half integer.
\subsection{Spinless Particles}
For $N$ spinless particles we distinguish two cases: N is an even or
an odd integer.

\textit{Case (1)}: If N is an even integer then
\begin{eqnarray}
I_{\mbox{even}}^{(0)}(\Delta)&=&\mbox{sgn}(\Delta)N(\frac{1}{2}-\Delta),
\: \cr
I_{\mbox{even}}^{(1)}(\Delta)&=&\alpha_{1}\alpha_{2}h(\Omega_{\tau_{1}})\sin(Nb_{\tau})\cos(1-2\Delta)b_{\tau},
\label{i01e}
\end{eqnarray}
where $\mbox{sgn}(x)$  equals +1, for $x>0$; 0 for $x=0$, and -1 for
$x<0$.
\textit{Case (2)}: If N is an odd integer then
\begin{eqnarray}
I_{\mbox{odd}}^{(0)}(\Delta)&=&-\mbox{sgn}(\Delta)\mbox{sgn}(\frac{1}{2}-\Delta)N(\frac{1}{2}-\bar\Delta),
\: \cr
I_{\mbox{odd}}^{(1)}(\Delta)&=&\alpha_{1}\alpha_{2}h(\Omega_{\tau_{1}})\sin(Nb_{\tau})\cos(1-2\bar\Delta)b_{\tau}.
\label{i01o}
\end{eqnarray}

\subsection{Particles with 1/2 spin}
For spin 1/2 particles we consider four cases.

\textit{Case (0)}: For an even number of particles' pairs, i.e. $N=4m$,
where $m$ is an integer we find
\begin{eqnarray}
I_{S}^{(0)}(\Delta_{S})=I_{\mbox{even}}^{(0)}(\Delta_{S}),\, 
I_{S}^{(1)}(\Delta_{S})=I_{\mbox{even}}^{(1)}(\Delta_{S}),
\label{is0}
\end{eqnarray}

\textit{Case (1)}: For an odd number of  particles' pairs, i.e. $N=4m+2$
we obtain
\begin{eqnarray}
I_{S}^{(0)}(\Delta_{S})=I_{\mbox{odd}}^{(0)}(\Delta_{S}), 
I_{S}^{(1)}(\Delta_{S})=I_{\mbox{odd}}^{(1)}(\Delta_{S}).
\label{is1}
\end{eqnarray}

\textit{Case (2)}: For an even number of pairs plus one extra particle,
i.e. $N=4m+1$ (there is one  particle whose spin is
unpaired as compared with case (0))  we find
\begin{eqnarray}
I_{\mbox{ext},S}^{(0)}(\Delta_{S})&=&-\mbox{sgn}(\Delta_{S})\mbox{sgn}(\frac{1}{2}-\Delta_{S})(\frac{N-1}{4}+\frac{1}{2}-\bar\Delta_{S}),
\: \cr
I_{\mbox{ext},S}^{(1)}(\Delta_{S})&=&\alpha_{1}\alpha_{2}h(\Omega_{\tau_{1}})\sin(b_{\tau})\cos(\frac{N-1}{2}+1-2\bar\Delta_{S})b_{\tau}.
\label{is2}
\end{eqnarray}
To determine which spin state is occupied by the extra particle one
compares the distance of the symmetric axis to the nearest half
integral axis, i.e.  $\bar\Delta_{S}$. The one with the larger
distance will be occupied.

\textit{Case (3)}: For an odd number of pairs plus one extra particle,
i.e. $N=4m+3$. Here we use case (1) and determine the
contribution to the current from the extra particle
\begin{eqnarray}
I_{\mbox{ext,S}}^{(0)}(\Delta_{S})&=&\mbox{sgn}(\Delta_{S})\mbox{sgn}(\frac{1}{2}-\Delta_{S})(\frac{N-3}{4}+\frac{1}{2}+\bar\Delta_{S}),
\: \cr
I_{\mbox{ext,S}}^{(1)}(\Delta_{S})&=&\alpha_{1}\alpha_{2}h(\Omega_{\tau_{1}})\sin(b_{\tau})\cos(\frac{N-3}{2}+1+2\bar\Delta_{S})b_{\tau}.
\label{is3}
\end{eqnarray}
Which spin state is occupied by the extra
particle is governed by $\bar\Delta_{S}$. The level with the smaller $\bar\Delta_{S}$
is populated.
\section{spin-dependent current (SC)}
In presence of a static magnetic field and the SOI but in the
absence of the pulse field the PCC is accompanied with a persistent
SC (PSC). Switching on the pulse field generates a spin-dependent
charge currents due to the SOI, and also a dynamic SC (DSC) that can
be controlled by the parameters of the pulse field.
The SC density  is
\[\mathbf{j}_{n_{0},S_{0}}^{s}(\mathbf{r}',t)=\Re\{\Psi_{n_{0}}^{S_{0},\dag}(\mathbf{r}',t)
\hat{\mathbf{v}}'\hat{\mathbf{s}}\Psi_{n_{0}}^{S_{0}}(\mathbf{r}',t)\},\]
 where
\[\hat{\mathbf{s}}=(\hbar/2)\hat{\sigma_{z}}\delta(\mathbf{r}'-\mathbf{r})\]
is the local spin density. The SC associated with level
$n_{0},S_{0}$ is
\begin{equation}
\mathbf{I}_{n_{0},S_{0}}^{s}(t>\tau)=\frac{1}{2\pi}\int_{0}^{2\pi}d
\varphi\int_{r_{1}}^{r_{2}}dr'\mathbf{j}_{n_{0},S_{0}}^{s}(\mathbf{r}',t)
\label{psc2}
\end{equation}
and can be evaluated as
\begin{equation}
\mathbf{I}_{n_{0},S_{0}}^{s_{z}}=\mathbf{I}_{s0}\Re\sum_{n}|C_{n}^{S_{0}}(n_{0},S_{0},t)|^{2}D_{1n}^{S_{0}},
\label{psc3}
\end{equation}
where \[\mathbf{I}_{s0}=\hat{\mathbf{e}}_{\varphi}E_{0}a/(2\pi)\]
 sets the unit  SC and
\begin{equation}
D_{1n}^{S_{0}}=[(a^{S_{0}})^{2}-(b^{S_{0}})^{2}](n-\frac{\phi}{\phi_{0}})-
(b^{S_{0}})^{2}. \label{d1n}
\end{equation}
 Here  \[(a^{S_{0}})^{2}-(b^{S_{0}})^{2}=S_{0}\cos\gamma,\] and
$S_{0}=\pm1$. The SC after
applying  two pulses to the ring is a sum of two
parts
\begin{equation}
\mathbf{I}_{n_{0},S_{0}}^{s_{z}}(t>\tau)=\mathbf{I}_{n_{0},S_{0}}^{s_{z},(0)}(t>\tau)+\mathbf{I}_{n_{0},S_{0}}^{s_{z},(1)}(t>\tau),
\label{psc4}
\end{equation}
where
\begin{eqnarray}
\mathbf{I}_{n_{0},S_{0}}^{s_{z},(0)}(t>\tau)&=&\mathbf{I}_{s0}[S_{0}\cos\gamma][(n-\frac{\phi}{\phi_{0}})+\frac{1}{2}-
\frac{S_{0}}{2\cos\gamma}], \: \cr
&=&
\mathbf{I}_{s0}[S_{0}\cos\gamma]\frac{I_{n_{0},S_{0}}^{(0)}(t>\tau)}{I_{0}},
\label{psct0}
\end{eqnarray}
is the static PSC \cite{sheng}
and the DSC part is
\begin{eqnarray}
\mathbf{I}_{n_{0},S_{0}}^{s_{z},(1)}(t>\tau)&=&\mathbf{I}_{s0}[S_{0}\cos\gamma][\alpha_{2}\langle\cos\varphi\rangle_{n_{0},S_{0}}(\tau)],
\: \cr
&=&
\mathbf{I}_{s0}[S_{0}\cos\gamma]\frac{I_{n_{0},S_{0}}^{(1)}(t>\tau)}{I_{0}}.
\label{psct1}
\end{eqnarray}
Summing over all  occupied energy levels we find
\begin{equation}
\mathbf{I}_{S_{0}}^{s_{z}}(t>\tau)=\mathbf{I}_{S_{0}}^{s_{z},(0)}(t>\tau)+\mathbf{I}_{S_{0}}^{s_{z},(1)}(t>\tau),
\label{tsc}
\end{equation}
where ($I_{S_{0}}^{(0),(1)}(t>\tau)$ are   PCC and  DCC)
\begin{equation}
\mathbf{I}_{S_{0}}^{s_{z},(0),(1)}(t>\tau)=\mathbf{I}_{s0}[S_{0}\cos\gamma]\frac{I_{S_{0}}^{(0),(1)}(t>\tau)}{I_{0}}.
\label{tsc1}
\end{equation}

\section{Numerical Results and Discussions}
We performed calculations for pulse-driven  ballistic quantum rings
fabricated by an appropriate confinement in a quantum well of
In$_{x}$Ga$_{1-x}$As/InP \cite{thsch}. Our results
 are also valid for other III-V or II-VI semiconductor
quantum rings with spin orbit, e.g. GaAs-AlGaAs quantum well,
or HgTe/HgCdTe quantum ring \cite{konig}. We shall present the total
charge current (TCC) which is a sum  of PCC and DCC over all
the occupied states.
Total spin-dependent current (TSC) is  obtained in the same way
\cite{splett}.
\begin{figure}[tbph]
\centering \includegraphics[width =10 cm, height=7 cm]{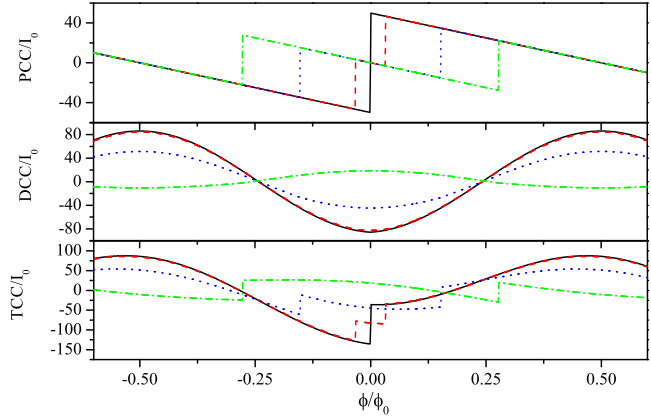}
\caption{(Color online) Persistent charge current (PCC), dynamic
charge current (DCC) and total charge current (TCC) are shown in
(a), (b) and (c) respectively. The spin orbit angle are
$\gamma=0^{\circ}, -20^{\circ}, -40^{\circ}$ and $-50^{\circ}$ for
the solid lines, the dash lines, the dot lines and the dash-dot
lines respectively in all graphs. The other parameters are $N=100$,
$a=100$ nm, $\tau=26.3$ ps, $F_{1}=F_{2}=1$ kV/cm.} \label{fig2}
\end{figure}

Fig. 2 shows how the flux and the SOI affect the PCC, DCC and TCC.
Without the SOI, the jump of the PCC occurs at integer flux for even
pair occupation, shown in Fig. 2. The jumps are different in other
occupations (see \cite{splett}), here we only focus on the even pair
occupation case for clarity. The periodic sawtooth  dependence of
the PCC on the flux exhibits has been studied before, e.g.
\cite{sheng,splett}. At finite SOI the jumps in PCC are shifted to
$\phi/\phi_{0}=l+(1\mp w)/2$; the two jumps are the consequences of
a superposition of the contributions from the two spin channels.
When the SOI strength is such that $\gamma=-\arccos(1/2n)$,
($n=1,2,\cdots$) the two jumps become at the half integer  which is
just the case of $4n+2$ occupation in absence of the SOI
\cite{splett}. The slope ratio between the two jumps is the same. As
can be inferred from the analytic expressions DCC (cf. Fig. 2)
depends smoothly on the flux. SOI results in a phase shift moving or
even exchanging the positions of the minima and maxima, as is for
$\gamma=-60^{\circ}$. The origin of the shape of TCC is deduced from
those of PCC and DCC. Here the magnitudes of the two contributions
is crucial: The PCC magnitude is related to the numbers of charge
carriers, while the DCC magnitude is determined primarily by the
product of the $\alpha_{1}$ and $\alpha_{2}$ (that can be externally
varied by changing the pulse intensities), the delay time $\tau$,
and the ring radius.
%

\begin{figure}[tbph]
\centering \includegraphics[width =11 cm, height=7 cm]{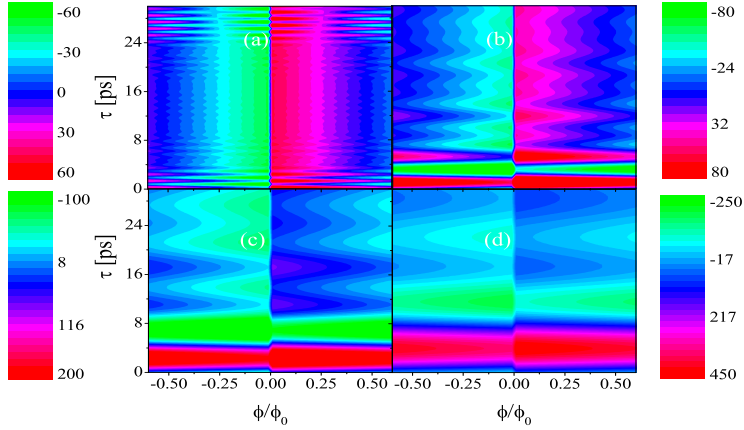}
\caption{(Color online) Contour plots of TCC on $\phi$ and $\tau$
are shown for different radius of the ring. $a=100$ nm, 200 nm, 300
nm and 400 nm in (a), (b), (c) and (d) respectively. The other
parameters are $N=100$, $\gamma=0^{\circ}$, $F_{1}=F_{2}=500$ V/cm.}
\label{fig3}
\end{figure}

Fig. 3 shows the TCC dependence on the ring radius (in the absence
of the SOI). As expected, a larger $\alpha$ enhances the DCC. On the
other hand, $\alpha$ enters the Bessel function argument whose
increase suppresses the magnitude of DCC. It can be shown that the
period of the oscillation with $\tau$ increases with increasing the
radius. The magnitudes of the maxima and minima are larger with
larger radius.

Now we  discuss the spin-dependent current  projected onto
the $z$ direction \cite{sheng}, i.e. $I^{S_{z}}$. The spin-dependent current projected onto the $\gamma$
direction (e.g. the quantization axis of the local spin frame) is
$I^{S_{z}}=I^{\gamma}\cos\gamma$ \cite{splett}.

\begin{figure}[tbph]
\centering \includegraphics[width =10 cm, height=7 cm]{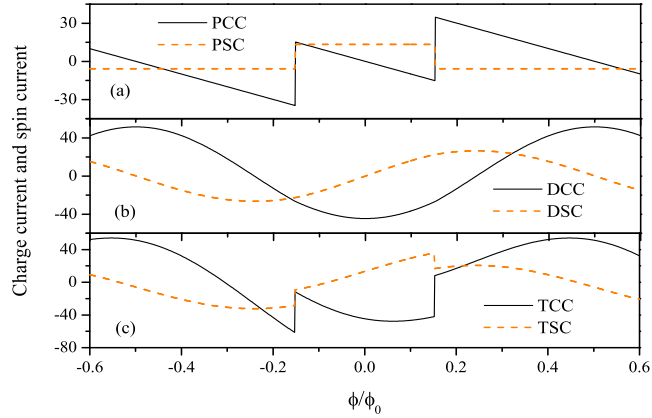}
\caption{(Color online) PCC and PSC, DCC and DSC, and TCC and TSC
are shown in (a), (b) and (c) respectively. The solid lines are the
charge current, and the dash lines are the spin current.
$\gamma=-40^{\circ}$, the other parameters are the same to Fig. 2.}
\label{fig4}
\end{figure}

PSC posses steps at PCC jumps (Fig. 4(a)) as a function of $\phi$.
This can be understood from the ratio of PCC for different spins;
the SOI only introduces a relative effective flux shift (see Eq.
(\ref{effflux})) leading to a constant spin-dependent current between the
jumps. For DSC vs. $\phi$ (Fig. 4(b))
 the effective flux leads to a shift of the DSC
along the flux axis. The physics behind this shift is that  SOI
provides a SU(2) flux, meaning that  the pulse-driven   (local
frame) spin-up electrons experience a different flux than those with
down spin, leading to a substantial spin-dependent current. In contrast, a
static magnetic flux does not induce a spin-dependent current in the absence
of the SOI. This shift and the jumps in the step function of the PSC
explain the behaviour of  TCC in Fig. 4(c).

\begin{figure}[tbph]
\centering \includegraphics[width =12 cm, height=7 cm]{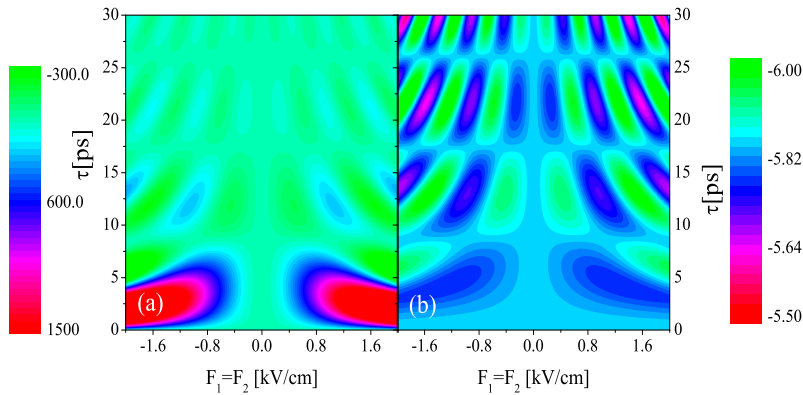}
\caption{(Color online) TCC (a) and TSC (b) vs.  pulses strengths
$F_1=F_2$ and delay time $\tau$. $\gamma=-40^{\circ}$, $\phi=0.2$.
Other parameters are $N=100$, $a=400$ nm.} \label{fig5}
\end{figure}

The control of TCC and TSC  by tuning the pulse-field parameters is
demonstrated in  Fig. 5. Because the  two pulses transfer a net
angular momentum setting the electrons in motion but they do not
couple directly to the spins the TCC and TSC show the same pattern
with $\tau$ and $F$.  From an experimental point of view it is
essential to note that we are dealing with non-equilibrium
quantities which opens the way for their detection via their
emission. E.g.,  the TCC and TSC can be detected by measuring the
current-induced magnetization of the ring  and the generated
electrostatic potential  \cite{scm}.

\section{General and concluding remarks}
In presence of spin-orbit coupling, two time-delayed appropriately
shaped electromagnetic pulses generate  spin-dependent charge
currents. As shown previously for the spin-independent case
\cite{matos05}, the sign  of the current and its magnitude  are
controllable via the  delay time and the strengths of the pulses.
From a symmetry viewpoint, similar phenomena may be expected to
occur for other geometries (wires, squares, etc).
 However, as shown for unbiased superlattices \cite{moskalenko1} (without SOI)
 details of the generated currents may differ qualitatively.
 Application of  an appropriate train of pulses open the possibility of
  controlling or even stopping  the current \cite{moskalenko}.
  For increasing the magnitude  of the current more intense pulses should be applied.

For generating currents in quantum rings one may also apply circular
polarized laser pulses \cite{pershin,rasanen}. In this context we
note the following: From an electrodynamics point of view,
generating  currents by our pulses is a completely classical effect,
i.e. currents are generated in a completely classical system, even
though in our case the subsequent excited carrier evolution is
quantum mechanical. For this reason our current is robust to
disorder and geometry modifications. In addition, the tunable time
delay between the pulses allows an ultra-fast control the current
properties. Using circular polarized laser pulses generates currents
for quantized systems (in which case the rotating-wave approximation
can be applied). For systems with  level broadening  on the order of
level spacing no appreciable current is generated. Our disadvantage
however is that our pulses are much more demanding to realize
experimentally, whereas laser pulses are readily available, in
particular at high light intensities allowing thus for a strong
current generation.

The DSC is proportional to the DCC which can be comparable to the
PCC for small or moderate occupation number case as seen in Fig.
(4b). The DCC depends  on the strength of the field and the delayed
time between the two pulses sensitively and dramatically. We provide
now an explicit calculation for the typical values of the CC and SC.
For In$_{x}$Ga$_{1-x}$As /InP quantum well \cite{thsch} we have
$m^{*}=0.037 m_{0}$. For a ring with radius of 100 nm, the line
velocity is then about $5000 m/s$, and the current unit is
$I_{0}/a\sim 8 nA$ which corresponds to the angular velocity current
for one particle. If we convert it into the unit of an induced
magnetization it is a radius-independent quantity $M_{0}\approx$2
meV/T (here we use the formula $M_{0}\approx\pi a^{2}(I_{0}/a)$
valid for rings considered here \cite{matos05}).

The work is support by the cluster of excellence "Nanostructured  Materials"
of the state Saxony-Anhalt.
%







\begin{thebibliography}{}
\bibitem{spintronics} For a review on spintronics, see, e.g., S. A. Wolf,
et al., Science \textbf{294}, 1488 (2001).

\bibitem{dresselhaus} G. Dresselhaus, Phys. Rev. \textbf{100}, 580
(1955).

\bibitem{rashba} E. I. Rashba, Sov. Phys. Solid State \textbf{2},
1109 (1960); Y. A. Bychkov, and E. I. Rashba, J. Phys. C
\textbf{17}, 6039 (1984).

\bibitem{lommer} G. Lommer, et al., Phys. Rev. Lett. \textbf{60},
728 (1988).

\bibitem{hgte} M. Schultz, et al., Semicond. Sci. Technol.
\textbf{11}, 1168 (1996); X. C. Zhang, et al., Phys. Rev. B
\textbf{63}, 245305 (2001); Y. S. Gui, et al., ibid. \textbf{70},
115328 (2004).


\bibitem{luo} J. Luo, et al., Phys. Rev. B \textbf{41}, 7685 (1990).

\bibitem{nitta} J. Nitta, et al., Phys. Rev. Lett. \textbf{78},
1335 (1997); C. -M. Hu, et al., ibid. \textbf{60}, 728 (1988); G.
Engels, et al., Phys. Rev. B \textbf{55}, R1958 (1997); Th.
Sch\"{a}pers, et al., J. Appl. Phys. \textbf{83}, 4324 (1998).

\bibitem{malcher} F. Malcher et al., Superlatt. Microstruc.
\textbf{2}, 267 (1986).




\bibitem{murakami} S. Murakami, N. Nagaosa, and S. C. Zhang, Science
\textbf{301}, 1348 (2003).

\bibitem{sinova} J. Sinova, et al., Phys. Rev. Lett. \textbf{92},
126603 (2004).

\bibitem{fuhrer} A. Fuhrer, et al., Nature \textbf{413}, 822 (2001);
Microelectronic Engineering \textbf{63}, 47 (2002); Phys. Rev. Lett.
\textbf{91}, 206802 (2003); ibid. \textbf{93}, 176803 (2004).

\bibitem{ring} R. J. Warburton, et al., Nature \textbf{405}, 926
(2000); A. Lorke, et al., Phys. Rev. Lett. \textbf{84}, 2223 (2000);
M. Bayer, et al., ibid. \textbf{90}, 186801 (2003); U. F. Keyser, et
al., ibid. \textbf{90}, 196601 (2003); B. Al\'{e}n, et al., Phys.
Rev. B \textbf{75}, 45319 (2007).

\bibitem{nitta99} J. Nitta, F. E. Meijer, and H. Takayanagi, Appl.
Phys. Lett. \textbf{75}, 695 (1999).

\bibitem{nitta03} J. Nitta, and T. Koga, J. Supercon. \textbf{16},
689 (2003).
\bibitem{hcp}  D. You,  R. R. Jones, P. H. Bucksbaum  and D. R. Dykaar, Opt.
Lett. \textbf{18}, 290 (1993);  T. J. Bensky,  G.  Haeffler, and R. R.  Jones, Phys. Rev.
Lett. \textbf{79}, 2018 (1997).

\bibitem{bennett} C. H. Bennett, and D. P. DiVincenzo, Nature
\textbf{404}, 247 (2000).

\bibitem{matos05} A. Matos-Abiague, and J. Berakdar, Phys. Rev. Lett.
\textbf{94}, 166801 (2005).

\bibitem{pershin} Y. V. Pershin, and C. Piermarocchi, Phys. Rev. B
\textbf{72}, 245331 (2005).

\bibitem{rasanen} E. R\"{a}s\"{a}nen, et al., Phys. Rev. Lett.
\textbf{98}, 157404 (2007).

\bibitem{zhu} Z.-G. Zhu, J. Berakdar, Phys. Rev. B \textbf{77},  235438
(2008).

\bibitem{ref1} J. Schliemann, J. C. Egues, and  Loss, Phys. Rev. Lett.
\textbf{90}, 146801 (2003); X. F. Wang  and P. Vasilopoulo, Phys.
Rev. B \textbf{72}, 165336 (2005).

\bibitem{matos} A. Matos-Abiague, and J. Berakdar, Phys. Rev. B
\textbf{70}, 195338 (2004); Europhys. Lett. \textbf{69}, 277 (2005).


\bibitem{meijer} F. E. Meijer, A. F. Morpurgo, and T. M. Klapwijk, Phys.
Rev. B \textbf{66}, 033107 (2002).

\bibitem{frustaglia} D. Frustaglia, and K. Richter, Phys. Rev. B
\textbf{69}, 235310 (2004).

%
\bibitem{molnar} B. Moln\'{a}r, \textit{et al.,}, Phys. Rev. B \textbf{69}, 155335 (2004); \textbf{72},
75330 (2005).

\bibitem{foldi} P. F\"{o}ldi, et al., Phys. Rev. B \textbf{71}, 33309 (2005); \textbf{73},
155325 (2006).

\bibitem{sheng} J. S. Sheng, and Kai Chang, Phys. Rev. B \textbf{74}, 235315
(2006).

\bibitem{splett} J. Splettstoesser, M. Governale, and U.
Z\"{u}licke, Phys. Rev. B \textbf{68}, 165341 (2003).

\bibitem{mizushima} M. Mizushima, Suppl. Prog. Theor. Phys. \textbf{40}, 207
(1967).

\bibitem{oh} S. Oh, and C. M. Ryu, Phys. Rev. B \textbf{51}, 13441
(1995).

\bibitem{frohlich} J. Fr\"{o}hlich, and U. M. Studer, Rev. Mod.
Phys. \textbf{65}, 733 (1993).

\bibitem{ballentine} (See Chap. 11 in) L. E. Ballentine, \textit{Quantum Mechanics: A Modern
Development}, World Scientific Publishing Co. Pte. Ltd, 1998.


\bibitem{loss} D. Loss, and P. Goldbart, Phys. Rev. B \textbf{43},
13762 (1991); ibid. \textbf{45}, 13544 (1992).

\bibitem{wendler} L. Wendler, and V. M. Fomin, A. A. Krokhin, Phys.
Rev. B \textbf{50}, 4642 (1994).

\bibitem{pccreview} For a review on quantum rings, see, e.g., A. G.
Aronov, and Yu. V. Sharvin, Rev. Mod. Phys. \textbf{59}, 755 (1987);
L. Wendler, and V. M. Fomin, Phys. Stat. Sol. (b) \textbf{191}, 409
(1995); U. Eckern, and P. Schwab, J. Low Tem. Phys. \textbf{126},
1291 (2002).

\bibitem{buttiker} M. B\"{u}ttiker, Y. Imry, and R. Landauer, Phys.
Lett. \textbf{96A}, 365 (1983).


\bibitem{chakraborty} T. Chakraborty, and P. Pietil\"{a}inen, Phys.
Rev. B \textbf{50}, 8460 (1994).


\bibitem{chandrasekhar} V. Chandrasekhar, et al., Phys. Rev. Lett. \textbf{67}, 3578
(1991).

\bibitem{mailly} D. Mailly, et al., Phys. Rev. Lett. \textbf{70},
2020 (1993).

\bibitem{sopcc}  Y. Meir, Y.
Gefen, and O. E.-Wohlman, Phys. Rev. Lett. \textbf{63}, 798 (1989);
O. E.-Wohlman, et al., Phys. Rev. B \textbf{45}, 11890 (1992).

\bibitem{moskalenko} A. S. Moskalenko, A. Matos-Abiague, and J.
Berakdar, Phys. Rev. B \textbf{74}, 161303 (2006); Europhys. Lett. \textbf{78}, 57001 (2007).

\bibitem{thsch} Th. Sch\"{a}pers, et al., J. Appl. Phys.
\textbf{83}, 4324 (1998).

\bibitem{konig} M K\"{o}nig, et al., Phys. Rev. Lett. \textbf{96},
76804 (2006).


\bibitem{scm} F. Meier and D. Loss, Phys. Rev. Lett. \textbf{90},
167204 (2003); F. Sch\"{u}tz, M. Kollar, and P. Kopietz, ibid.
\textbf{91}, 017205 (2003); Q.-F. Sun, et al., cond-mat/0301402; F.
S. Nogueira, and K. -H. Bennemann, cond-mat/0302528.

\bibitem{moskalenko1} A. S. Moskalenko, A. Matos-Abiague, and J. Berakdar, Phys. Lett. A \textbf{356}, 255
(2006).




\end{thebibliography}
\end{document}